\documentclass[runningheads]{llncs}

\usepackage[utf8]{inputenc} 
\usepackage[T1]{fontenc}    
\usepackage{hyperref}       
\usepackage{url}            
\usepackage[anythingbreaks]{breakurl}

\usepackage{breakurl}
\usepackage{hyperref}
\usepackage{booktabs}       
\usepackage{amsfonts}       
\usepackage{nicefrac}       
\usepackage{microtype}      
\usepackage{lipsum}
\usepackage{graphicx}
\usepackage{amsmath}
\usepackage[skip=1pt, indent=40pt]{parskip}
\usepackage[table]{xcolor}
\usepackage[square,numbers]{natbib}
\title{\bf COBRA: Comparison-Optimal Betting for Risk-limiting Audits}

\author{Jacob V. Spertus}

\institute{
University of California, Berkeley, 
Department of Statistics\\
\email{jakespertus@berkeley.edu}
}

\authorrunning{
J.V. Spertus
} 

\date{\today}

\begin{document}

\maketitle

\begin{abstract}
    Risk-limiting audits (RLAs) can provide routine, affirmative evidence that reported election outcomes are correct by checking a random sample of cast ballots.
    An \textit{efficient} RLA requires checking relatively few ballots.
    Here we construct highly efficient RLAs by optimizing supermartingale tuning parameters---\textit{bets}---for ballot-level comparison audits. 
    The exactly optimal bets depend on the true rate of errors in cast-vote records (CVRs)---digital receipts detailing how machines tabulated each ballot.
    We evaluate theoretical and simulated workloads for audits of contests with a range of diluted margins and CVR error rates. 
    Compared to bets recommended in past work, using these optimal bets can dramatically reduce expected workloads---by 93\% on average over our simulated audits. 
    Because the exactly optimal bets are unknown in practice, we offer some strategies for approximating them.
    As with the ballot-polling RLAs described in ALPHA and RiLACs, adapting bets to previously sampled data or diversifying them over a range of suspected error rates can lead to substantially more efficient audits than fixing bets to \textit{a priori} values, especially when those values are far from correct. 
    We sketch extensions to other designs and social choice functions, and conclude with some recommendations for real-world comparison audits.
\end{abstract}

\keywords{risk-limiting audit, election integrity, comparison audit, nonparametric testing, betting martingale}

\section{Introduction}

Machines count votes in most American elections, and (reported) election winners are declared on the basis of these machine tallies. 
Voting machines are vulnerable to bugs and deliberate malfeasance, which may undermine public trust in the accuracy of reported election results. 
To counter this threat, risk-limiting audits (RLAs) can provide routine, statistically rigorous evidence that reported election outcomes are correct---that reported winners really won---by manually checking a demonstrably secure ballot trail \citep{lindemanStark12}. 
RLAs have a user-specified maximum chance---the \textit{risk limit}---of certifying a wrong reported outcome, and will never overturn a correct reported outcome.
They can also be significantly more \textit{efficient} than full hand counts, requiring fewer manually tabulations to verify a correct reported outcome and reducing costs to jurisdictions. 

There are various ways to design RLAs. 
Data can be sampled as batches of ballots (i.e. precincts or machines) or as individual ballot cards (hereafter, we refer to cards simply as ``ballots").
Sampling individual ballots is more statistically efficient than sampling batches.
In a \textit{polling audit}, sampled ballots are checked directly without reference to machine interpretations. 
Ballot-polling audits sample and check individual ballots.
In a \textit{comparison audit}, manual interpretations of ballots are compared to their machine interpretations. 
Ballot-level comparison audits check each sampled ballot against a corresponding \textit{cast vote record} (CVR)---a digital receipt detailing how the machine tallied the ballot. 
Not all voting machines can produce CVRs, but ballot-level comparison audits are the most efficient type of RLA. 

Ideally, RLAs will simultaneously check multiple (potentially all) contests within a jurisdiction---a task made considerably more efficient by targeting samples with card-style data \citep{GlazerEtal21}. Card-style data are most feasibly derived from CVRs, in which case each contest can be audited using ballot-level comparison. Because the overall workload of the audit is aggregated across contests, optimizing the efficiency for individual contests can provide substantial workload reductions for the audit as a whole. Thus, constructing sharper ballot-level comparison audits is paramount to the implementation of real-world RLAs auditing multiple contests.

The earliest RLAs were formulated for batch-level comparison audits, which are analogous to historical, statutory audits \citep{stark08a}. Subsequently, the maximum across contest relative overstatement (MACRO) was used for comparison RLAs \citep{stark09b, stark09d, stark10d, ottoboniEtal18}, but its efficiency suffered from conservatively pooling observed errors across candidates and contest.
SHANGRLA \citep{stark20a} unified RLAs as hypotheses about means of lists of bounded numbers and provided sharper methods for batch and ballot-level comparisons. 
Each null hypothesis tested in a SHANGRLA-style RLA posits that the mean of a bounded list of \textit{assorters} is less than 1/2.
If all the nulls are declared false at risk limit $\alpha$, the audit can stop.
Any valid test for the mean of a bounded finite population can be used to test these hypotheses, allowing RLAs to use a wide range of \textit{risk-measuring functions}. 

Betting supermartingales (BSMs)---described in \citet{waudby-smithEtal21} and \citet{stark22}---provide a particularly useful class of risk-measuring functions.
BSMs are \textit{sequentially valid}, allowing auditors to update and check the measured risk after each sampled ballot while maintaining the risk limit.
They can be seen as generalizations of risk-measuring functions used in earlier RLAs, including Kaplan-Markov, Kaplan-Kolmogorov, and related methods \citep{stark09b, stark20a}.
They have tuning parameters $\lambda_i$ called \textit{bets}, which play an important role in determining the efficiency of the RLA.
Previous papers using BSMs for RLAs have focused on setting $\lambda_i$ for efficient ballot-polling audits; betting for comparison audits has been treated as essentially analogous \citep{waudby-smithEtal21, stark22}.
However, as we will show, comparison audits are efficient with much larger bets than are optimal for ballot-polling.

This paper details how to set BSM bets $\lambda_i$ for efficient ballot-level comparison audits, focusing on audits of plurality contests.
Section \ref{sec:notation} reviews SHANGRLA notation and the use of BSMs as risk-measuring functions. 
Section \ref{sec:oracle_bets} derives optimal ``oracle'' bets under the Kelly criterion \citep{kelly56}, which assumes knowledge of true error rates in the CVRs. 
In reality, these error rates are unknown, but the oracle bets are useful in constructing practical betting strategies, which plug in estimates of the true rates.
Section \ref{sec:practical_betting} presents three such strategies: guessing the error rates \textit{a priori}, using past data to estimate the rates adaptively, or positing a distribution of likely rates and diversifying bets over that distribution. 
Section \ref{sec:simulations} presents two simulation studies: one comparing the oracle strategy derived in \citet{waudby-smithEtal21} for ballot-polling against our comparison-optimal strategy, and one comparing practical strategies against one another. 
Section \ref{sec:extensions} sketches some extensions to betting while sampling without replacement and to social choice functions beyond plurality.
Section \ref{sec:conclusion} concludes with a brief discussion and recommendations for practice.

\section{Notation}
\label{sec:notation}

\subsection{Population and parameters} 
Following SHANGRLA \citep{stark20a} notation, let $\{c_i\}_{i=1}^N$ denote the CVRs, $\{b_i\}_{i=1}^N$ denote the true ballots, and $A()$ be an \textit{assorter} mapping CVRs or ballots into $[0,u]$. 
We will assume we are auditing a plurality contest, in which case $u := 1$, $A(b_i) := 1$ if the ballot shows a vote for the reported winner, $A(b_i) := 1/2$ if it shows an undervote or vote for a candidate not currently under audit, and $A(b_i) := 0$ if it shows a vote for the reported loser. 
The \textit{overstatement} for ballot $i$ is $\omega_i := A(c_i) - A(b_i)$.
$\bar{A}^c := N^{-1} \sum_{i=1}^N A(c_i)$ is the average of the assorters computed on the CVRs. 
Finally, the comparison audit population is comprised of \textit{overstatement assorters}
$x_i := (1 - \omega_i) / (2 - v)$, 
where $v := 2 \bar{A}^c - 1$ is the \textit{diluted margin}: the difference in votes for the reported winner and reported loser, divided by the total number of ballots cast. 

Let $\bar{x} := N^{-1} \sum_{i=1}^N x_i$ be the average of the comparison audit population and $\bar{A}^b := N^{-1} \sum_{i=1}^N A(b_i)$ be the average of the assorters applied to ballots. 
Section 3.2 of \citet{stark20a} establishes the relations 
$$ \mbox{reported outcome is correct} \iff \bar{A}^b > 1/2 \iff \bar{x} > 1/2.$$
As a result, rejecting the \textit{complementary null} 
\begin{equation}
    H_0: \bar{x} \leq 1/2 \label{eqn:comp_null}
\end{equation}
at risk limit $\alpha$ provides strong evidence that the reported outcome is correct. 

Throughout this paper, we ignore \textit{understatement} errors---those in favor of the reported winner with $\omega_i < 0$.
Understatements help the audit end sooner, but will generally have little effect on the optimal bets.
We comment on this choice further in Section \ref{sec:conclusion}.
With this simplification, overstatement assorters comprise a list of numbers $\{x_i\}_{i=1}^N \in \{0, a/2, a\}^N$ where $a := (2-v)^{-1} > 1/2$ corresponds to the value on correct CVRs, $a/2$ corresponds to 1-vote overstatements, and 0 corresponds to 2-vote overstatements.
This population is parameterized by 3 fractions: 
\begin{itemize}
    \item $p_0 := \#\{x_i = a\} / N$ is the rate of correct CVRs. 
    \item $p_1 := \#\{x_i = a/2\}/N$ is the rate of 1-vote overstatements.
    \item $p_2 := \#\{x_i = 0\} / N$ is the rate of 2-vote overstatements.
\end{itemize} 
The population mean can be written $\bar{x} = a p_0 + (a/2) p_1$.

\subsection{Audit data}
Ballots may be drawn by sequential simple random sampling with or without replacement, but we first focus on the with replacement case for simplicity. 
Implications for sampling without replacement are discussed in Section \ref{sec:extensions}. 
We have a sequence of samples $X_1,X_2,\ldots \overset{\mbox{\tiny iid}}{\sim} F$, where $F$ is a three-point distribution with mass $p_0$ at $a$, $p_1$ at $a/2$, and $p_2$ at $0$. 

\subsection{Risk measurement via betting supermartingales}
Let $T_i := 1 + \lambda_i (X_i - 1/2)$ where $\lambda_i \in [0,2]$ is a freely-chosen tuning parameter that may depend on past samples $X_1,\ldots,X_{i-1}$. 
Define $M_0 := 1$ and
$$M_t := \prod_{i=1}^t T_i = \prod_{i=1}^t [1 + \lambda_i (X_i - 1/2)].$$
$M_t$ is a \textit{betting supermartingale} (BSM) for any \textit{bets} $\lambda_i \in [0,2]$ whenever (\ref{eqn:comp_null}) holds because 
$$\bar{x} \leq 1/2 \implies \mathbb{E}[X_i \mid X_{i-1},...,X_1] \leq 1/2 \implies \mathbb{E}[M_t \mid X_{t-1},\ldots,X_1] \leq M_{t-1}  $$ 
where the first implication comes from simple random sampling with replacement. 

Intuitively, $M_t$ can be thought of as the wealth accumulated by a gambler who starts with 1 unit of capital at time $t=0$ and at time $t=i$ stakes proportion $\lambda_i$ of their current capital on observing $X_i > 1/2$. If $\lambda_i = 0$, they stake nothing and can neither gain nor lose capital on round $i$. If $\lambda_i = 2$, they stake everything and can lose all their capital if $X_i = 0$. For any bets that depend only on past data, the gambler cannot expect to accumulate wealth by betting that $X_i > 1/2$ when (\ref{eqn:comp_null}) is true.
Ville's inequality \citep{ville39} then guarantees that it is unlikely that the gambler's wealth ever becomes large: 
$$\mathbb{P}(\exists ~ t \in \mathbb{N} : M_t \geq 1/\alpha) \leq \alpha.$$
For example, when (\ref{eqn:comp_null}) holds, the probability that the gambler ever accumulates more than 20 units of wealth is no more than 0.05.

As a matter of risk measurement, Ville's inequality implies that the truncated reciprocal $P_t := \min\{1, 1/M_t\}$ is a sequentially-valid $P$-value for the complementary null in the sense that
$\mathbb{P}(\exists ~ t \in \mathbb{N} : P_t \leq \alpha) \leq \alpha$
when $\bar{x} \leq 1/2$ for any risk limit $\alpha \in (0,1)$.
More details on BSMs are given in \citet{WaudbysmithRamdas20, waudby-smithEtal21} and \citet{stark22}.
To obtain an efficient RLA, we would like to make $M_t$ as large as possible ($P_t$ as small as possible) when $\bar{x} > 1/2$. 

\section{Oracle betting}
\label{sec:oracle_bets}
We begin by deriving ``oracle" bets by assuming we can access the true error rates $p_0$, $p_1$, and $p_2$ and optimizing the expected growth of the logarithm of the martingale under these rates. We call these oracle bets because they are exactly optimal for this objective, but depend on unknown parameters and hence cannot be implemented in practice. However, oracle bets can be approximated to run efficient comparison audits with the practical betting strategies discussed in Section \ref{sec:practical_betting}.

\subsection{Error-free CVRs}

\label{sec:no_cvr_error}

\begin{figure}[h]
    \centering
    \includegraphics[width = \textwidth]{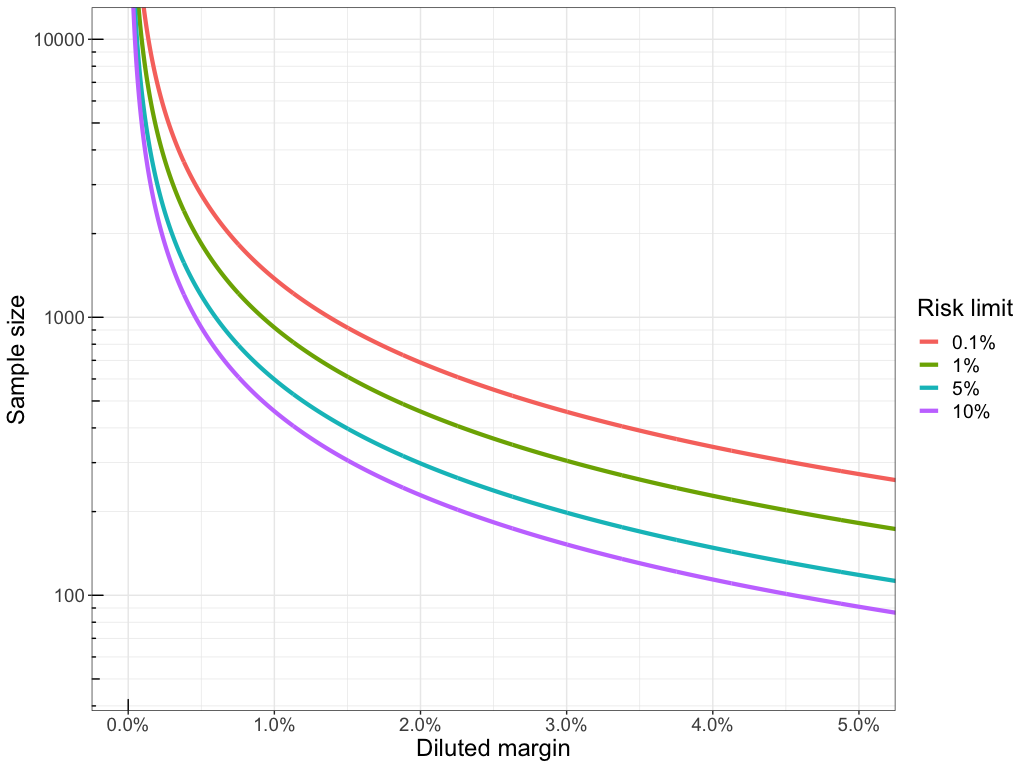}
    \caption{Deterministic sample sizes (y-axis; $\log_{10}$ scale) for a comparison audit of a plurality contest with various diluted margins (x-axis) and risk limits (colors), with no error in CVRs and a maximal bet of $\lambda = 2$ on every draw.}
    \label{fig:minimum_sample_sizes}
\end{figure}

In the simple case where there is no error at all in the CVRs, $p_0 = 1$ and $x_i = \bar{x} = a$ for all $i$. 
When computing the BSM, it doesn't matter which ballot is drawn:
$$T_i = 1 + \lambda_i (a - 1/2) \mbox{ and } M_t = [1 + \lambda_i (a - 1/2)]^t.$$ 
Because $(a-1/2) > 0$, the best strategy is to bet as aggressively as possible, setting $\lambda_i := 2$.  Under such a bet, $M_t = (2a)^t$. 
Setting this equal to $1/\alpha$ yields the stopping time:
\begin{equation}
    t_{\tiny{\mbox{stop}}} = \frac{\log (1/\alpha)}{\log (2 a)} = \frac{-\log(\alpha)}{\log(2) - \log(2 - v)} \label{eqn:no_error_lb}
\end{equation}
where $v$ is the diluted margin. 
Ignoring understatement errors, (\ref{eqn:no_error_lb}) is a deterministic lower bound on the sample size of a comparison audit when risk is measured by a BSM.
Figure \ref{fig:minimum_sample_sizes} plots this as a function of the diluted margin and risk limit. 

\subsection{Betting with CVR Error}
\label{sec:cvr_error}
Usually CVRs will have at least some errors, and maximal bets are far from ideal when they do. We now show why this is true before deriving an alternative oracle strategy. 
In general,
$$T_i = \begin{cases} 1 + \lambda_i (a - 1/2) &\mbox{with probability}~ p_0\\ 1 + \lambda_i(a/2 - 1/2) &\mbox{with probability}~ p_1\\
1 - \lambda_i/2 &\mbox{with probability}~ p_2.\end{cases}$$
If we fix $\lambda_i := \lambda$ and try to maximize $M_n$ by maximizing the expected value of each $T_i$, we find
$\mathbb{E}_F[T_i] = p_0 [1 + \lambda(a - 1/2)] + p_1 [1-\lambda (1-a)/2] + p_2[1 - \lambda/2]
    = 1 + (a p_0 + \frac{a}{2} p_1 - 1/2) \lambda.$
This is linear with a positive coefficient on $\lambda$, since $a p_0 + \frac{a}{2} p_1 = \bar{x} > 1/2$ under any alternative. 
Therefore, the best strategy seems to be to set $\lambda := 2$ as before.
However, unless $p_2 = 0$, $M_t$ will eventually ``go broke" with probability 1: $T_i = 0$ if a 0 is drawn while the bet is maximal. Then $M_t = 0$ for all future times and we cannot reject at any risk limit $\alpha$. In this case, we say the audit \textit{stalls}: it must proceed to a full hand count to confirm the reported winner really won.

To avoid stalls we follow the approach of \citet{kelly56}, instead maximizing the expected value of $\log T_i$.
The derivative is
\begin{equation}
    \frac{d}{d \lambda} \mathbb{E}_F[\log T_i] = \frac{(a-1/2)p_0}{1 + \lambda(a-1/2)} + \frac{(a-1)p_1}{2 - \lambda (1-a)} - \frac{p_2}{2 - \lambda}. \label{eqn:derivative}
\end{equation}
The oracle bet $\lambda^*$ can be found by setting this equal to 0 and solving for $\lambda$ using a root-finding algorithm. 

Alternatively, we can find a simple analytical solution by assuming no 1-vote overstatements and setting $p_1 = 0$. In this case, solving for $\lambda$ yields:
\begin{equation}
    \lambda^* = \frac{2 - 4 a p_0}{1 - 2a} \label{eqn:optimal_lambda}
\end{equation}
Note that $\lambda^* > 0$ since $a p_0 > 1/2$ under the alternative, and $\lambda^* < 2$ since $a > 1/2$. 


\subsection{Relation to ALPHA}
\label{sec:alpha}
There is a one-to-one correspondence between oracle bets for the BSM $M_t$ and oracle bets for the ALPHA supermartingale, which reparameterizes $M_t$. 
Note that the list of overstatement assorters $\{x_i\}_{i=1}^N$ is upper bounded by the value of a 2-vote understatement,
$u := 2 / (2-v)  = 2 a$. 
Section 2.3 of \citet{stark22} shows that the equivalently optimal $\eta$ for use with ALPHA is:
$$\eta^* :=  1/2(1 + \lambda^*(u - 1/2)) = \frac{1 - 2 a p_0}{2-4a} + 2a p_0 - 1/2 .$$
Naturally, when $p_0=1$, $\eta^* = 2a = u$, which is the maximum value allowed for $\eta^*$ while maintaining ALPHA as a non-negative supermartingale.

\section{Betting in Practice}
\label{sec:practical_betting}

In practice, we have to estimate the unknown overstatement rates to set bets. 
We propose and evaluate three strategies: fixed, adaptive, and diversified betting. 
Throughout this section, we use $\tilde{p}_k$ to denote a generic estimate of $p_k$ for $k \in \{1,2\}$. 
When the estimate adapts in time, we use the double subscript $\tilde{p}_{ki}$. 
In all cases, the estimated overstatement rates are ultimately plugged into (\ref{eqn:derivative}) to estimate the optimal bets.

\subsection{Fixed betting}

The simplest approach is to make a fixed, \textit{a priori} guess at $p_k$ using historic data, machine specifications, or other information. 
For example, $\tilde{p}_1 := 0.1\%$ and $\tilde{p}_2 := 0.01\%$ will prevent stalls and may perform reasonably well when there are few overstatement error. 
This strategy is analagous to apKelly for ballot-polling, which fixes $\lambda_i$ based on an \textit{a priori} estimate of the population assorter mean (typically derived from reported tallies). However, \citet{waudby-smithEtal21} and \citet{stark22} show that apKelly can become quite poor when the estimate is far from correct.
This frailty motivates more sophisticated strategies. 

\subsection{Adaptive betting}

In a BSM, the bets need not be fixed and $\lambda_i$ can be a \textit{predictable} function of the data $X_1,\ldots,X_{i-1}$, since we condition on these data when establishing $M_t$ as a martingale. 
Intuitively, the gambler can adapt their bets based on outcomes of previous rounds and, if the null is true, still cannot expect to gain capital in the next round. 
This fact allows us to estimate error rates based on past samples in addition to $\textit{a priori}$ considerations when setting $\lambda_i$. We adapt the ``truncated-shrinkage'' estimator introduced in Section 2.5.2 of \citet{stark22} to rate estimation. For $k \in \{1,2\}$ we set a value $d_k \geq 0$, capturing the degree of shrinkage to the \textit{a priori} estimate $\tilde{p}_k$, and a truncation factor $\epsilon_k \geq 0$, enforcing a lower bound on the estimated rate.
Let $\hat{p}_{ki}$ be the sample rates at time $i$, e.g., $\hat{p}_{2i} = i^{-1} \sum_{j=1}^i 1\{X_j = 0\}$.
Then the truncated-shrinkage estimate is:
\begin{equation}
    \tilde{p}_{ki} := \frac{d_k \tilde{p}_k + i \hat{p}_{k(i-1)}}{d_k + i - 1} \vee \epsilon_k \label{eqn:shrink_trunc}
\end{equation}
The rates are allowed to learn from past data in the current audit through $\hat{p}_{k(i-1)}$, while being anchored to the \textit{a priori} estimate $\tilde{p}_k$. The tuning parameter $d_k$ reflects the degree of confidence in the \textit{a priori} rate, with large $d_k$ anchoring more strongly to $\tilde{p}_k$. Finally, $\epsilon_k$ should generally be set above 0 to prevent stalls.  

At each time $i$, the truncated-shrinkage estimated rate $\tilde{p}_{ki}$ can be plugged into (\ref{eqn:derivative}) and set equal to 0 to obtain the bet $\lambda_i$. Fixing $\tilde{p}_{1i} := 0$ allows us to use (\ref{eqn:optimal_lambda}), in which case $\lambda_i = (2 - 4 a (1-\tilde{p}_{2i})) / (1-2a)$.

\subsection{Diversified betting}

A weighted average of BSMs:
$$\sum_{b=1}^{B} \theta_b \prod_{i=1}^t [1 + \lambda_b (X_i - 1/2)],$$
where $\theta_b \geq 0$ and $\sum_{b=1}^B \theta_b = 1$, is itself a BSM. 
The intuition is that our initial capital is split up into $B$ pots, each with $\theta_b$ units of wealth. 
We then bet $\lambda_b$ on each pot at each time, and take the sum of the winnings across all pots as our total wealth at time $t$. 
\citet{WaudbysmithRamdas20} construct the ``grid Kelly" martingale by defining $\lambda_b$ along an equally spaced grid on $[0,2]$ and giving each the weight $\theta_b = 1/B$. \citet{waudby-smithEtal21} refine this approach into ``square Kelly'' for ballot-polling RLAs by placing more weight at close margins.

We adapt these ideas to the comparison audit context by parameterizing a discrete grid of weights for $p_1$ and $p_2$. 
We first note that $(p_1,p_2)$ are jointly constrained by the hyperplane $a p_2 + (a/2) p_1 \leq a - 1/2$ under the alternative, since otherwise there is enough error to overturn the reported result.
A joint grid for $(p_1, p_2)$ can be set up by separately constructing two equally-spaced grids from 0 to $v/k$, computing the Cartesian product of the grids, and removing points where $a p_2 + (a/2) p_1 \geq a - 1/2$.  
Once a suitable grid has been constructed, the weights at each point can be flexibly defined to reflect the suspected rates of overstatements. 
At each point $(p_1,p_2)$, $\lambda_b$ is computed by passing the rates $(p_1,p_2)$ into (\ref{eqn:derivative}) and solving numerically; the weight for $\lambda_b$ is $\theta_b$.
Thus a distribution of weights on the grid of overstatement rates induces a distribution on the bets.

\begin{figure}[h]
    \centering
    \includegraphics[width = \textwidth]{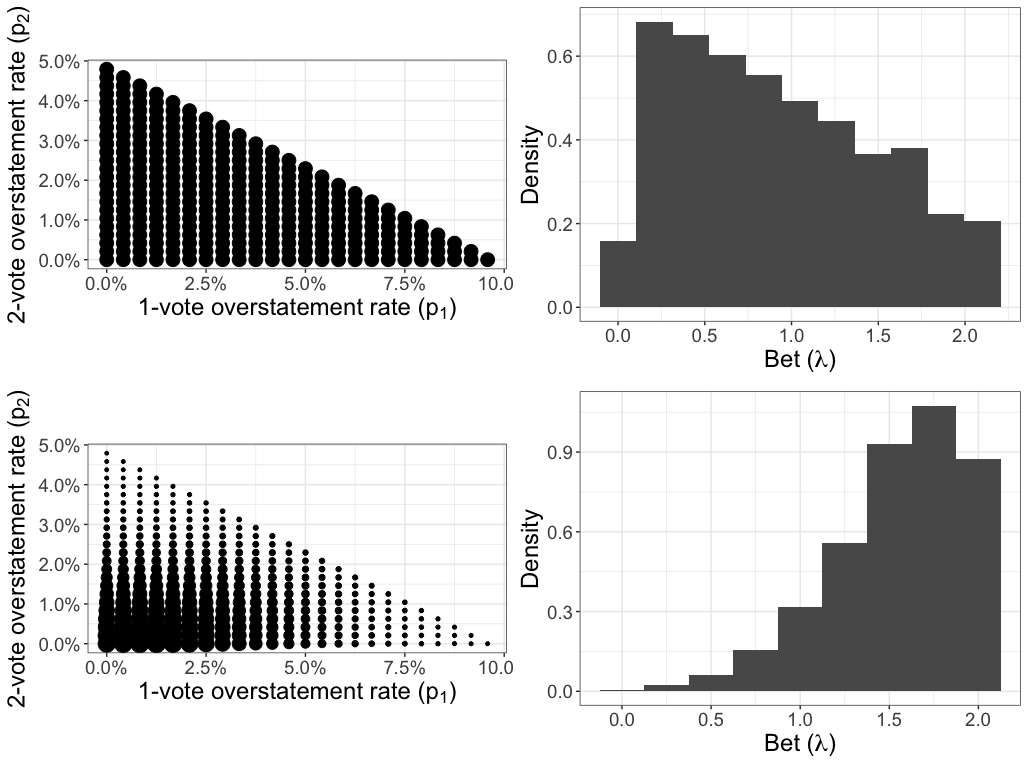}
    \caption{Plots showing two mixture distributions over overstatement rates (left column; y-axis = 2-vote overstatment rate, x-axis = 1-vote overstatement rate; point size = mixture weight) and their corresponding induced distributions over the bets (right column; x-axis = bet, y-axis = density).
    The diluted margin of 10\% constrains possible overstatement rates.
    The upper row shows a uniform grid of weights over all overstatement rates (left column) and its induced distribution on $\lambda$ (right column). The bottom row plots discretized, truncated, and rescaled bivariate normal weights with parameters $(\mu_1, \mu_2) = (.01, .001)$, $(\sigma_1, \sigma_2) = (.02, .01)$, and $\rho = 0.25$ (left column) and its induced distribution on $\lambda$ (right column).}
    \label{fig:overstatment_plots}
\end{figure}

Figure \ref{fig:overstatment_plots} illustrates two possible weighted grids for a diluted margin of $v=10\%$, and their induced distribution on bets $\{\lambda_b\}_{b=1}^B$.
In the top row, the weights are uniform with $\theta_b = 1/B$. 
In the bottom row, the weights follow a bivariate normal density with mean vector and covariance matrix respectively specified to capture a prior guess at $(p_1, p_2)$ along with the uncertainty in that guess. 
The density is truncated, discretized, and rescaled so that the weights sum to unity.

\section{Numerical evaluations}
\label{sec:simulations}

We conducted two simulation studies.
The first evaluated stopping times for bets using the oracle comparison bets in (\ref{eqn:optimal_lambda}) against the oracle value of apKelly from \citet{waudby-smithEtal21}. 
The second compared stopping times for oracle bets and the 3 practical strategies we proposed in Section \ref{sec:practical_betting}. 
All simulations were run in R (version 4.1.2).

\subsection{Oracle simulations}

We evaluated stopping times of oracle bets at multiple diluted margins and 2-vote overstatement rates when sampling with replacement from a population of size $N = 10000$. 
At each combination of diluted margin $v \in \{0.05, 0.10, 0.20\}$ and 2-vote overstatement rates $p_2 \in \{1.5\%, 1\%, 0.5\%, 0.1\%, 0\%\}$ we ran 400 simulated comparison audits.
We set $p_1 = 0$: no 1-vote overstatements.

The bets corresponded to oracle bets $\lambda^*$ in Equation (\ref{eqn:optimal_lambda}) or to $\lambda^{\tiny{\mbox{apK}}} := 4\bar{x} - 2$, the ``oracle'' value of the apKelly strategy in Section 3.1 of \citet{waudby-smithEtal21} and Section 2.5 of \citet{stark22}\footnote{$\lambda^{\tiny{\mbox{apK}}}$ implies a bet of $\eta_i := \bar{x}$ in the ALPHA parameterization.}, which were originally derived for ballot-polling.
$\lambda^{\tiny{\mbox{apK}}}$ uses the true population mean instead of an estimate based on reported tallies. 
In each scenario, we estimated the expected and 90th percentile workload from the empirical mean and 0.9 quantile of the stopping times at risk limit $\alpha = 5\%$ over the 400 simulations. To compare the betting strategies, we computed the ratios of the expected stopping time for $\lambda^*$ over $\lambda^{\tiny{\mbox{apK}}}$ in each scenario. We then took the geometric mean across scenarios as the average reduction in expected workload.

Table \ref{tab:stopping_times} presents the mean and 90th percentile (in parentheses) stopping times over the 400 simulations. BSM comparison audits with $\lambda^*$ typically require counting fewer than 1000 ballots, and fewer than 100 for wide margins without CVR errors. 
On average, betting by $\lambda^*$ provides an enormous advantage over $\lambda^{\tiny{\mbox{apK}}}$: the geometric mean workload ratio is 0.072, a 93\% reduction.

\begin{table}[ht]
\centering
\begin{tabular}{ll|ll}
  
  & & \multicolumn{2}{c}{\textbf{Stopping times}}\\
  \cline{3-4}
 \textbf{DM} & \textbf{2-vote OR} & $ \textbf{apKelly}~ (\lambda^{\tiny{\mbox{apK}}}$) & \textbf{Oracle} ($\lambda^*$) \\ 
  \hline
   5\% & 1.5\% & 10000 (10000) & 1283 (2398) \\ 
   & 1.0\% & 10000 (10000) & 482 (813) \\ 
   & 0.5\% & 7154 (7516) & 242 (389) \\ 
   & 0.1\% & 4946 (5072) & 146 (257) \\ 
   & 0.0\% & 4559 (4559) & 119 (119) \\ 
   \hline
   10\% & 1.5\% & 2233 (2464) & 177 (323) \\ 
   & 1.0\% & 1705 (1844) & 131 (233) \\ 
   & 0.5\% & 1346 (1429) & 83 (116) \\ 
   & 0.1\% & 1130 (1167) & 65 (60) \\ 
   & 0.0\% & 1083 (1083) & 59 (59) \\ 
   \hline
   20\% & 1.5\% & 339 (371) & 52 (78) \\ 
   & 1.0\% & 304 (335) & 42 (57) \\ 
   & 0.5\% & 272 (289) & 35 (61) \\ 
   & 0.5\% & 249 (258) & 30 (29) \\ 
   & 0.0\% & 245 (245) & 29 (29) \\ 
   \hline
\end{tabular}
\label{tab:stopping_times}
\caption{Mean (90th percentile) stopping times of 400 simulated comparison audits run with oracle bets ($\lambda^*$) or apKelly bets ($\lambda^{\tiny{\mbox{apK}}}$) under a range of diluted margins and 2-vote overstatement rates. DM = diluted margin; OR = overstatement rate.}
\end{table}
\subsection{Practical simulations}
\label{sec:practical_comparisons}

We evaluated oracle betting, fixed \textit{a priori} betting, adaptive betting, and diversified betting in simulated comparison audits with $N=20000$ ballots, a diluted margin of 5\%, 1-vote overstatement rates $p_1 \in \{0.1\%, 1\%\}$, and 2-vote overstatement rates $p_2 \in \{0.01\%, 0.1\%, 1\%\}$. 

Oracle bets were set using the true values of $p_1$ and $p_2$ in each scenario. 
The other methods used prior guesses $\tilde{p}_1 \in \{0.1\%, 1\%\}$ and $\tilde{p}_2 \in \{0.01\%, 0.1\%\}$ as tuning parameters in different ways.
The fixed method derived the optimal bet by plugging in $\tilde{p}_k$ as a fixed value.
The adaptive method anchored the truncated-shrinkage estimate $\tilde{p}_{ki}$ displayed in equation (\ref{eqn:shrink_trunc}) to $\tilde{p}_k$, but updated using past data in the sample. 
The tuning parameters were $d_1 := 100$, $d_2 := 1000$, $\epsilon_1 = \epsilon_2 := 0.001\%$. 
The larger value for $d_2$ reflects the fact that very low rates (expected for 2-vote overstatements) are harder to estimate empirically, so the prior should play a larger role. 
The diversified method used $\tilde{p}_k$ to set the mode of a mixing distribution, as in the lower panels of Figure \ref{fig:overstatment_plots}.
Specifically, the mixing distribution was a discretized, truncated, bivariate normal with mean vector $(\tilde{p}_1, \tilde{p}_2)$, standard deviation $(\sigma_1, \sigma_2) := (0.5\%, 0.25\%)$, and correlation $\rho := 0.25$. 
The fact that $\sigma_2 < \sigma_1$ reflects more prior confidence that 2-vote overstatement rates will be concentrated near their prior mean, while $\rho > 0$ encodes a prior suspicion that overstatement rates 
are correlated: they are more likely to be both high or both low. 
After setting the weights at each grid point according to this normal density, they were rescaled to sum to unity. 

We simulated 400 audits under sampling with replacement for each scenario. 
The stopping times were capped at 20000, the size of the population, even if the audit hadn't stopped by that point.
We estimated the expected value and 90th percentile of the stopping times for each method by the empirical mean and 0.9 quantile over the 400 simulations. 
We computed the geometric mean ratio of the expected stopping times of each method over that of the oracle strategy as a summary of their performance across scenarios.

Table \ref{tab:table_2} presents results. With few 2-vote overstatements, all strategies performed relatively well and the audits concluded quickly. 
When the priors substantially underestimated the true overstatement rates, the performance of the audits degraded significantly compared to the oracle bets. 
This was especially true for the fixed strategy. For example, when $(p_1, p_2) = (0.1\%, 1\%)$ and $\tilde{p}_2 = 0.01\%$, the expected number of ballots for the fixed strategy to stop was more than 20 times that of the oracle method. 
On the other hand, the adaptive and diversified strategies were much more robust to a poor prior estimate.
In particular, the expected stopping time of the diversified method was never more than 3 worse than that of the oracle strategy, and the adaptive method was never more than 4 times worse. 
The geometric mean workload ratios of each strategy over the oracle strategy were 2.4 for fixed, 1.3 for adaptive, and 1.2 for diversified. 
The diversified method was the best practical method on average across scenarios. 

\begin{table}[ht]
\centering
\scalebox{.9}{
\begin{tabular}{r|r|rr|llll}
  \multicolumn{2}{c|}{\textbf{True ORs}} & \multicolumn{2}{c|}{\textbf{Prior ORs}} & \multicolumn{4}{c}{\textbf{Stopping Times}} \\ 
  \hline
$p_2$ & $p_1$ & $\tilde{p}_2$ & $\tilde{p}_1$ & Oracle & Fixed & Adaptive & Diversified \\ 
  \hline
0.01\% & 0.1\% & 0.01\% & 0.1\% & 124 (147) & 125 (119) & 124 (147) & 131 (152) \\ 
  &  &  & 1\% & 124 (147) & 125 (147) & 125 (147) & 131 (154) \\
  
   &  & 0.1\% & 0.1\% & 125 (147) & 129 (151) & 131 (151) & 133 (155) \\ 
   &  &  & 1\% & 127 (147) & 132 (153) & 130 (152) & 135 (157) \\ 
   \cline{2-8}
   & 1\% & 0.01\% & 0.1\% & 174 (229) & 167 (229) & 166 (229) & 177 (236) \\ 
   &  &  & 1\% & 168 (229) & 172 (229) & 167 (229) & 180 (235) \\ 
   
   &  & 0.1\% & 0.1\% & 176 (229) & 169 (232) & 175 (262) & 181 (262) \\ 
   &  &  & 1\% & 159 (205) & 174 (233) & 180 (265) & 184 (264) \\ 
   \hline
  0.1\% & 0.1\% & 0.01\% & 0.1\% & 146 (256) & 153 (338) & 159 (350) & 149 (271) \\ 
   &  & & 1\% & 151 (256) & 154 (174) & 150 (147) & 145 (154) \\ 
   
   &  & 0.1\% & 0.1\% & 147 (256) & 152 (256) & 146 (182) & 153 (259) \\ 
   &  &  & 1\% & 149 (256) & 151 (244) & 147 (256) & 152 (265) \\ 
   \cline{2-8}
   & 1\% & 0.01\% & 0.1\% & 209 (351) & 227 (420) & 225 (460) & 214 (400) \\ 
   &  &  & 1\% & 200 (324) & 240 (457) & 232 (500) & 211 (378) \\ 
   
   &  & 0.1\% & 0.1\% & 204 (351) & 208 (364) & 210 (358) & 208 (344) \\ 
   &  &  & 1\% & 208 (324) & 205 (324) & 205 (341) & 219 (371) \\ 
   \hline
  1\% & 0.1\% & 0.01\% & 0.1\% & 526 (996) & 13654 (20000) & 1581 (3517) & 888 (2090) \\ 
   & &  & 1\% & 525 (984) & 12685 (20000) & 1585 (3731) & 739 (1708) \\
   
  &  & 0.1\% & 0.1\% & 528 (1032) & 9589 (20000) & 1112 (2710) & 812 (1982) \\ 
   & &  & 1\% & 534 (985) & 7247 (20000) & 915 (2294) & 686 (1586) \\ 
   \cline{2-8}
  & 1\% & 0.01\% & 0.1\% & 999 (1908) & 15205 (20000) & 3855 (7811) & 2637 (5873) \\ 
  &  &  & 1\% & 1110 (2002) & 15641 (20000) & 3477 (7529) & 1803 (4331) \\ 
  
   & & 0.1\% & 0.1\% & 1030 (1868) & 13113 (20000) & 2795 (5996) & 2064 (4884) \\ 
   &  &  & 1\% & 1127 (2256) & 13094 (20000) & 2437 (5452) & 1604 (3758) \\ 
   \hline
\end{tabular}
}
\caption{Mean (90th percentile) stopping times over 400 simulated comparison audits with diluted margin of $v = 5\%$ and varying overstatement rates at risk limit $\alpha = 5\%$. The true overstatement rates are in the first two columns. The second two columns contain the prior guesses of the true overstatement rates, used to set bets differently in each strategy as described in Section \ref{sec:practical_comparisons}. The oracle strategy uses the true rates to set the bets, so all variation over $(\tilde{p}_1$, $\tilde{p}_2)$ in the results for that strategy is Monte Carlo variation. Monte Carlo variation also accounts for any differences in the fixed and oracle strategies when $(\tilde{p}_1, \tilde{p}_2) = (p_1, p_2)$, since the bets are identical. Note that some stopping time distributions are highly skewed, e.g. the 90th percentile is lower than the mean for fixed bets with $(\tilde{p}_1, \tilde{p}_2) = (p_1, p_2) = (0.1\%, 0.01\%)$. OR = overstatement rate.}
\label{tab:table_2}
\end{table}
\section{Extensions}
\label{sec:extensions}

\subsection{Betting while sampling without replacement}

When sampling without replacement, the distribution of $X_i$ depends on past data $X_1,...,X_{i-1}$. 
Naively updating an \textit{a priori} bet to reflect what we know has been sampled may actually harm the efficiency of the audit.

Specifically, recall that, for $k \in \{1,2\}$, $\hat{p}_{ki}$ denotes the sample proportion of the overstatement rate at time $i$. If we fix initial rate estimates to $\tilde{p}_{k}$, then the updated estimate at time $i$ given that we have removed $i \hat{p}_{k (i-1)}$ would be
$$\tilde{p}_{ki} = \frac{N \tilde{p}_k - i \hat{p}_{k(i-1)}}{N - i + 1} ~~\mbox{for}~~ k \in \{1,2\}.$$
This can be plugged into (\ref{eqn:derivative}) to estimate the optimal $\lambda_i^*$ for each draw.
Fixing $\tilde{p}_{1i} = 0$ and using equation (\ref{eqn:optimal_lambda}) yields the closed form optimum:
$$\lambda_i^* = \frac{2 - 4 a \tilde{p}_{2i}}{1 - 2a} \wedge 2,$$
where we have truncated at 2 to guarantee that $\lambda_i^*$ is even a valid bet. 
This is necessary because the number of 2-vote overstatements in the sample can exceed the number $N\tilde{p}_2$ hypothesized to be in the entire population. 
If this occurs, the audit will stall if even one more 2-vote overstatement is discovered.
More generally, this strategy has the counterintuitive (and counterproductive) property of betting \textit{more} aggressively as more overstatements are discovered.
To avoid this pitfall we suggest using the betting strategies we derived earlier under IID sampling, even when sampling without replacement. 


\subsection{Other social choice functions}

SHANGRLA \citep{stark20a} encompasses a broad range of social choice functions beyond plurality, all of which are amenable to comparison audits. 
Assorters for approval voting and proportional representation are identical to plurality assorters, so no modification to the optimal bets is required. 
Ranked-choice voting can also be reduced to auditing a collection of plurality assertions, though this reduction may not be the most efficient possible \citep{blom2019}.
On the other hand, some social choice functions, including weighted additive and supermajority, require different assorters and will have different optimal bets. 

In a supermajority contest, the diluted margin $v$ is computed differently depending on the fraction $f \in (1/2, 1]$ required to win, as well as the proportion of votes for the reported winner in the CVRs. In the population of overstatement assorters error-free CVRs still appear as $a = (2-v)^{-1}$, but 2-vote overstatements are $(1 - 1/(2f)) a > 0$ and 1-vote overstatements are $(3/2 - 1/(2f)) a$. 
So that the population attains a lower bound of 0, we can make the shift $x_i - (1 - 1/(2f)) a$ and test against the shifted mean $1/2 - (1-1/(2f)) a$. 
Because there are only 3 points of support, the derivations in 
Section \ref{sec:cvr_error} can be repeated, yielding a new solution for $\lambda^*$ in terms of the rates and the shifted mean. 

Weighted additive schemes apply an affine transformation to ballot scores to construct assorters. 
Because scores may be arbitrary non-negative numbers, there can be more than 3 points of support for the overstatement assorters and the derivations in Section \ref{sec:cvr_error} cannot be immediately adapted. 
If most CVRs are correct then most values in the population will be above 1/2, suggesting that an aggressive betting strategy with $\lambda := 2-\epsilon$ will be relatively efficient. 
Alternatively, a diversified strategy weighted towards large values of $\lambda \in (0,2]$ can retain efficiency when there are in fact high rates of error.
It should also be possible to attain a more refined solution by generalizing the optimization strategy in Section \ref{sec:cvr_error} to populations with more than 3 points of support.

\subsection{Batch-level comparison audits}

Batch-level comparison audits check for error in totals across batches of ballots, and are applicable in different situations than ballot-level comparisons, since they do not require CVRs.
SHANGRLA-style overstatement assorters for batch-level comparison audits are derived in \citet{stark22}.
These assorters generally take a wide range of values within $[0,u]$.
Because they are not limited to a few points of support, there is not a simple optimal betting strategy. 
However, assuming there is relatively little error in the reported batch-level counts, will again place the majority of the assorter distribution above 1/2.
This suggests using a relatively aggressive betting strategy, placing more weight on bets near 2 (or near the assorter upper bound in the ALPHA parameterization).

\citet{stark22} evaluated various BSMs in simulations approximating batch-level comparison audits, though the majority of mass was either at 1 or spread uniformly on $[0,1]$, not at a value $a \in (1/2,1]$. Nevertheless, in situations where most of the mass was at 1, aggressive betting ($\eta \geq 0.9$) was most efficient. Investigating efficient betting strategies for batch-level comparison audits remains an important area for future work.

\section{Conclusions}
\label{sec:conclusion}

We derived optimal bets for ballot-level comparison audits of plurality contests and sketched some extensions to broader classes of comparison RLAs. 
The high-level upshot is that comparison should use considerably more aggressive betting strategies than polling in practice, a point made abundantly clear in our oracle simulations. 
Our practical strategies approached the efficiency of oracle bets, except in cases where $p_2 = 1\%$. 
Such a high rate of 2-vote overstatements is unlikely in practice, and would generally imply something has gone terribly wrong: votes for the loser should not be flipped to votes for the winner. 

Future work should continue to flesh out efficient strategies for batch-level comparison, and explore the effects of understatement errors. 
We suspect that understatements will have little effect on the optimal strategy. If anything, they imply bets should be even \textit{more} aggressive, but we already suggest placing most weight near the maximal value of $\lambda_i = 2$, diversifying or thresholding to prevent stalls if 2-vote overstatements are discovered. 
We hope our results will guide efficient real-world comparison RLAs, and demonstrate the practicality of their routine implementation for trustworthy, evidence-based elections.
\\

\section*{Acknowledgements}

Philip Stark and Amanda Glazer provided helpful feedback on an earlier draft of this paper. Jacob Spertus' research is supported by NSF grant 2228884. 

\section*{Code}
Code implementing our simulations and generating our figures and tables is available on Github at \url{https://github.com/spertus/comparison-RLA-betting}.

\footnotesize
\bibliography{pbsBib.bib}

\end{document}